\documentclass[preprintnumbers,aps,amsmath,superscriptaddress,nofootinbib,tightenlines,twocolumn]{revtex4}
\usepackage{graphicx}
\usepackage{bm}
\usepackage{epsfig}

\def\msb{{\overline{\rm MS}}}
\def\lqcd{\Lambda_{\text{QCD}}}
\def\mbos{m_b^{\rm 1S}}

\def\nn{\nonumber \\ }
\def\mcmc{ \overline m_c\left( \overline m_c \right) }
\def\mcmu{ \overline m_c\left( \mu \right) }
\def\mcmc{ \overline m_c\left( \overline m_c \right) }
\def\mcmum{ \overline m_c\left( \mu_m \right) }
\def\mcmumnf{ \overline m_c^{(n_f=4)}\left(\mu_m\right) }
\def\norder{ \varphi }

\begin{document}


\preprint{ \vbox{ 
\hbox{MPP-2005-100}
\hbox{UCSD-PTH-05-13} 
}}

\title{Charm Quark Mass from Inclusive Semileptonic B Decays}

\author{Andr\'e~H.~Hoang}
\affiliation{Max-Planck-Institut f\"ur Physik (Werner-Heisenberg-Institut), 
F\"ohringer Ring 6, 80805 M\"unchen, Germany}

\author{Aneesh~V.~Manohar}

\affiliation{Max-Planck-Institut f\"ur Physik (Werner-Heisenberg-Institut), 
F\"ohringer Ring 6, 80805 M\"unchen, Germany}
\affiliation{University of California, San Diego,
9500 Gilman Drive, La Jolla, CA 92093}


\begin{abstract}
\vspace{0.5cm}
The $\msb$ charm quark mass is determined to be $\mcmc=1224 \pm 17 \pm 54$~MeV from
a global fit to inclusive $B$ meson decay data, where the first error is
experimental, and includes the uncertainty in $\alpha_s$,  and the second is
an estimate of theoretical uncertainties in
the computation. We discuss the implications of the pole mass renormalon
in the determination of $m_c$. 
\end{abstract}
\maketitle


\newpage

%
%
%
\section{Introduction}
\label{sectionintroduction}

Precise determinations of the bottom and charm quark masses are becoming
increasingly important for the measurement of CKM parameters and in 
the search for new physics from experiments on
B-meson and Kaon decays. For example, the theoretical uncertainty in the $K^+
\to \pi^+ \nu \bar \nu$ rate is now dominated by the uncertainty in 
$\mcmc$~\cite{Buras:2005gr}. At present, the most precise determinations of
the charm quark mass have been obtained from sum rule analyses of the $e^+e^-$
R-ratio and from lattice QCD (though all lattice results still use
the quenched approximation)~\cite{Brambilla:2004wf}. 
The results of recent analyses are
consistent~\cite{Kuhn:2001dm,Hoang:2004xm,Rolf:2002gu,deDivitiis:2003iy}; 
nevertheless it is important to consider other independent methods to
extract the charm quark mass. 
Recently, fundamental standard model parameters such as the bottom quark mass
and the mixing angle $V_{cb}$ have been determined with high precision from
inclusive semileptonic
$B$-decays~\cite{Bauer:2002sh,Bauer:2004ve,Gambino:2004qm,Aubert:2004aw}. In
this paper, we extend the inclusive $B$ decay 
analysis to also obtain the $\msb$ charm quark mass, $\mcmc$. The charm quark
mass was extracted previously in  Ref.~\cite{Bauer:2004ve} by the methods
discussed here. We redo the analysis including a more careful   
consideration of renormalon effects in the quark pole masses. 

The experimental input consists of inclusive $B$ decay spectra, such as the
electron energy and hadronic invariant mass spectra in inclusive semileptonic
$B \to X_c e \overline \nu$ decay, the inclusive photon energy spectrum in $B
\to X_s \gamma$ decay, and the $B$ meson masses and lifetimes. These
quantities can be computed theoretically using heavy quark effective theory
(HQET), as an expansion in $\alpha_s(m_b)$ and $\lqcd/m_b$. At present, all
the quantities are known to order $\alpha_s^2 \beta_0$, $\lqcd^3/m_b^3$, and
$\alpha_s\lqcd/m_b$, and fits to the experimental data were performed in
Ref.~\cite{Bauer:2004ve} including all theoretical quantities to this
order. We use the results of this fit to determine the charm quark mass.

It is often convenient to do computations using the quark pole masses
$m_{b,c}$ since they define the phase space boundaries in partonic
perturbation theory. The heavy quark pole masses are known to have a renormalon
ambiguity of order $\lqcd$~\cite{bigi,beneke}. The order $\lqcd$ renormalon means that the pole
masses $m_{b,c}$ cannot be determined accurately (i.e.\ with an uncertainty
much smaller than $\lqcd$), 
and the $\alpha_s$ perturbation expansion for the pole masses is poorly
behaved, with a factorial growth in the coefficients at high orders. The bad
behavior of the pole masses is already manifest in the first two orders of
perturbation theory.  
In Ref.~\cite{Bauer:2002sh}, fits to $B$ decay spectra in terms of the pole
mass were shown to be poorly behaved. Luckily, there are other quark mass
definitions which do not suffer from a renormalon ambiguity, and which can be
determined accurately. One of these, the 1S-mass for the bottom quark, $\mbos$,
was used as the bottom quark mass definition in the fits of
Refs.~\cite{Bauer:2002sh,Bauer:2004ve}. It was shown in
Ref.~\cite{Bauer:2002sh} that fits to inclusive $B$ decays using the 1S-mass
are well-behaved, with good convergence of the perturbation series. 
Similarly, for the charm quark, a quark mass definitions without the $\lqcd$
renormalon is the $\msb$ mass $\mcmu$, frequently given for
$\mu=\overline m_c$, $\mcmc$.

The difference between the bottom and charm quark masses is constrained by the
meson mass difference, 
\begin{eqnarray}
\overline m_B - \overline m_D  &=& m_b^{\rm pole}-m_c^{\rm pole} - {\lambda_1 \over 2}
\left({1\over m_b } - {1 \over m_c } \right) \nn
&&+ { \rho_1 - \tau_1 - \tau_3 \over 4}\left({1\over m_b^2 } - {1 \over m_c^2 } \right) 
\label{mbmcdiff}
\end{eqnarray}
where $\overline m_{B,D}$ are the spin-averaged $B,B^*$ and $D,D^*$ meson
masses, and $\lambda_1, \rho_1, \tau_{1,3}$ are HQET parameters. This equation
was used to determine the bottom-charm pole mass difference $m_b-m_c=3.41\pm
0.01$~GeV in Ref.~\cite{Bauer:2004ve}, where we frequently suppress the
superscript ``pole'' for the pole masses. The order $\lqcd$ renormalon cancels in
bottom-charm pole mass difference, so $m_b-m_c$ has no $\lqcd$
renormalon, and can be accurately determined. 

The procedure for accurately determining the charm quark mass is simple in
principle --- one combines  the values of $m_b-m_c$ and $\mbos$ to determine
$\mcmu$. None of the three quantities suffers from a $\lqcd$ renormalon, and
so $\mcmu$ can 
be determined reliably. However, it turns out there are some subtleties in the
actual computation, and one must take care to ensure that the pole mass
renormalon cancellation is properly implemented in $m_b-m_c$ at intermediate
stages of the calculation. This was not done correctly in
Ref.~\cite{Bauer:2004ve}. Including the renormalon cancellation properly, as
done here, leads to significantly different results.

\section{Perturbative Mass Relations }
\label{sectionrelations}

The computations of the moments of the semileptonic decay spectra are
carried out in the theory where the top quark is integrated out and
where the bottom quark is treated as static. Thus the top and bottom
quarks do not contribute in virtual quark loops, and all $\msb$ parameters are
defined in the theory for $n_f=4$ running flavors. 

The 3-loop perturbative series for the relation between the charm pole
and the charm $\msb$ mass at the scale $\mu_m$ is~\cite{Melnikov:2000qh}
\begin{eqnarray}
&& m_c^{\rm pole}  = 
\mcmumnf\ g\left(\alpha_s\left(\mu\right),\mu,\mcmumnf,\mu_m\right), \nonumber\\[10pt]
&& g(\alpha_s,\mu,\overline m,\mu_m)  = 
1+ \epsilon \, \delta^{(1)}(\alpha_s,\mu,\overline m,\mu_m)\nonumber\\[5pt]
&&\hspace{1cm}+\epsilon^2 \, \delta^{(2)}(\alpha_s,\mu,\overline m,\mu_m)
+\epsilon^3 \, \delta^{(3)}(\alpha_s,\mu,\overline m,\mu_m)
\label{mcpolemsbar} 
\end{eqnarray}
where $\alpha_s\equiv \alpha_s^{(n_f=4)}(\mu)$ is the running coupling in the
theory with four dynamical flavors, $\overline m\equiv \overline 
m_c^{(n_f=4)}(\mu_m)$ is the $\msb$ charm mass in the theory with four
dynamical flavors, and $\delta^{(i)}(\alpha_s,\mu,\overline m,\mu_m)$ are: 
\begin{eqnarray}
\delta^{(1)} & = & 
  \frac{\alpha_s}{\pi}\,\left( \frac{4}{3} + L_m\right)\,,
\\[3mm]
\delta^{(2)} & = & 
  \frac{\alpha_s^2}{\pi^2}\,\Big(10.3193 + 2.77778\,L + \nonumber\\[2mm]
&&       \left(2.98611 + 2.08333\, L\right)\,L_m - 
        0.541667\, L_m^2 \Big)\,,
\\[3mm] 
\delta^{(3)} & = & 
  \frac{\alpha_s^3}{\pi^3}\,\Big(116.504 + 47.2748\,L + 5.78704\,L^2 + \nonumber\\[2mm]
&&    \left( 6.30337 + 15.6505\,L + 4.34028\,L^2\right)\,L_m
\nonumber\\[2mm] & &  + 
    \left(-6.20023 - 2.25694\,L\right)\,L_m^2 +
     0.571759\,L_m^3\Big)\nn
\end{eqnarray}
where
\begin{eqnarray}
L_m  \equiv  \ln\Big(\frac{\mu_m^2}{\overline m^2}\Big),\qquad
L  \equiv  \ln\Big(\frac{\mu^2}{\overline m^2}\Big)
\,.
\end{eqnarray}
For most of the paper, the superscript $(n_f=4)$ for $\msb$ parameters
is dropped for simplicity.
The powers of the auxiliary parameter $\epsilon=1$ indicate the
respective orders in the loop expansion. Note that the charm quark mass and
$\alpha_s$ are renormalized at the scale $\mu_m$ and  $\mu$,
respectively. Choosing 
$\mu=\mu_m=\mcmc$ and $\alpha_s(\overline m_c)=0.39$ gives 
\begin{eqnarray}
g &=& 1 + 0.42 \alpha_s(\overline m_c) \epsilon + 1.05 
 \alpha_s(\overline m_c)^2 \epsilon^2 +  3.76 \alpha_s(\overline m_c)^3 \epsilon^3 \nn
 &=& 1 + 0.17 \epsilon + 0.16 \epsilon^2 + 0.22 \epsilon^3.
 \label{7}
\end{eqnarray}
The first few terms of this series show no sign of convergence, and seem to
indicate that one cannot determine $\mcmc$ to an accuracy better than, say,
20\%. The poor perturbative behavior of Eq.~(\ref{7}) is a reflection of the
${\cal O}(\Lambda_{\rm QCD})$ pole mass renormalon, which implies that the
series for the pole mass behaves at high orders as  $\sim \mu
\epsilon^\norder\alpha_s(\mu)^\norder \norder! \beta_0^\norder$, where
$\norder$ is the order in perturbation theory, 
$\beta_0=11-2/3 n_l$ is the one-loop coefficient of the QCD $\beta$ function,
$n_l$ is the number of light quark flavors, and $\mu$ the renormalization
scale for $\alpha_s$. As we will see explicitly in the next section, the bad
behavior manifest in Eq.~(\ref{7}) does not limit the accuracy with
which one can determine $\mcmu$ because in the relation between physical 
observables and short-distance masses, such as $\mcmu$ and $\mbos$,  the 
${\cal O}(\Lambda_{\rm QCD})$ renormalon contributions always completely
cancel. To achieve this cancellation in practical numerical analyses a few 
subtle guidelines for the treatment of truncated perturbative series have
to be followed.  

For the series between the bottom quark pole and the bottom 1S masses
the higher order corrections each contain a factor of the inverse 
$\Upsilon$ Bohr radius $\sim \mbos \alpha_s$ in addition to the powers
of the strong coupling. For the systematic cancellation of the 
${\cal O}(\Lambda_{\rm QCD})$ renormalon contributions in each order
of the perturbative series one has to use the upsilon
expansion~\cite{Hoang:1998ng,Hoang:1998hm}, i.e.\ terms of order
$\alpha_s^{n+1}$ in the perturbative series between the bottom quark
pole and the bottom 1S masses are formally treated of order
$\alpha_s^n$. For this reason, the powers of $\epsilon$ and $\alpha_s$ differ
by one in Eq.~(\ref{mbpoleMb1S}). From now on, we will use $\epsilon$ as the
perturbation series expansion parameter, and the order to which we expand will
be denoted by $\norder$. 

The third order relation between the bottom pole and 1S
masses reads~\cite{Hoang:1999ye,Hoang:2000fm}
($\alpha_s\equiv \alpha_s^{(n_f=4)}(\mu), \overline m\equiv \overline
m_c^{(n_f=4)}(\mu_m)$)
\begin{eqnarray}
&& m_b^{\rm pole}  =  \mbos f(\mbos,\alpha_s(\mu),\mu,\overline m(\mu_m),\mu_m)
\nonumber \\[10pt]
&& f(\mbos,\alpha_s,\mu,\overline m,\mu_m)  =   1
+ \epsilon\,\bigg[\,
    \Delta^{(1)}(\mbos,\alpha_s)  \,\bigg]
\nonumber
\\[2mm]
& & 
+ \, \epsilon^2\,\bigg[\,
\Delta^{(2)}
(\mbos,\alpha_s,\mu)+
\Delta^{(2)}_{\mbox{\tiny m}}(\mbos,\alpha_s,\mu,\overline m)
  \,\bigg]
\nonumber
\\[2mm]
& & 
+ \, \epsilon^3\,\bigg[\,
\Delta^{(3)}(\mbos,\alpha_s,\mu)+
\Delta^{(3)}_{\mbox{\tiny m}}(\mbos,\alpha_s,\mu,\overline m,\mu_m)
  \,\bigg]\nn
\label{mbpoleMb1S}
\end{eqnarray}
where 
\begin{eqnarray}
\Delta^{(1)} & = &
\frac{2}{9}\alpha_s^2 \,,
\end{eqnarray}
\begin{eqnarray}
\Delta^{(2)} & = &
\frac{2}{9}\frac{\alpha_s^3}{\pi}\Big( 11.2778 + 8.33333\, L_{\rm 1S} \Big)
+ (\Delta^{(1)})^2
\,,
\label{Deltamassless2}
\end{eqnarray}
\begin{eqnarray}
\Delta^{(3)} & = &  
\frac{2}{9}\frac{\alpha_s^4}{\pi^2}\Big( 
 192.693 + 119.083\,L_{\rm 1S} + 52.0833\,L_{\rm 1S}^2\Big)
\nonumber \\[2mm] & &
+\, 2\,\Delta^{(1)}\,\Delta^{(2)} 
+ (\Delta^{(1)})^3 
- \frac{\alpha_s\beta_0}{\pi}\,(\Delta^{(1)})^2
\,,
\label{Deltamassless3}
\end{eqnarray}
\begin{eqnarray}
L_{\rm 1S} & \equiv & \ln\Big(\frac{3\,\mu}{4\,\alpha_s M_b^{\rm
    1S}}\Big)
\,,
\end{eqnarray}
and
\begin{eqnarray}
\Delta^{(2)}_{\mbox{\tiny m}} 
& = &
\Delta^{\mbox{\tiny NLO}}_{\mbox{\tiny massive}}(\overline m,\mbos,\alpha_s)
\,,
\label{Deltamassive2}
\\[4mm]
\Delta^{(3)}_{\mbox{\tiny m}} 
& = &
\Delta^{\mbox{\tiny NNLO}}_{\mbox{\tiny m}}(\overline m,\mbos,\alpha_s,\mu)
+ 2 \Delta^{(1)}\,\Delta^{\mbox{\tiny NLO}}_{\mbox{\tiny m}}(\overline
m,\mbos,\alpha_s)
\nonumber\\[2mm] & &
+\,  \Big[\delta^{(1)}-\Delta^{(1)}\Big]\,
\overline\Delta^{\mbox{\tiny NLO}}_{\mbox{\tiny m}}(\overline m,\mbos,\alpha_s,\mu)
\,.
\label{Deltamassive3}
\end{eqnarray}
The latter two terms describe the corrections arising from the
non-zero charm quark mass. 
These contributions are included for
consistency, since the charm quark is  treated as massive 
in the charm pole-$\msb$ mass relation of Eq.~(\ref{mcpolemsbar}).
Note that accounting for the charm mass corrections in
Eq.~(\ref{mbpoleMb1S}) is mandatory to ensure the complete
cancellation of the pole mass renormalon contributions in
$m_b-m_c$~\cite{Hoang:2000fm}. This is 
because, in the perturbation series, only quarks that are treated as massless
contribute to the bad higher order behavior that leads to the order
$\Lambda_{\rm QCD}$ renormalon ambiguity. To be more specific, only massless
quarks contribute to the number of light quarks $n_l$ that governs the
asymptotic large order  behavior; light quarks
whose masses are not set to zero in the computations, as well as heavy quarks that
are integrated out or treated as static, do not contribute to the
$\Lambda_{\rm QCD}$ renormalon ambiguity in the perturbative
series since their masses effectively act as an infrared cutoff for
perturbative contributions from small momenta. The charm quark has been
treated as massive in Eq.~(\ref{mcpolemsbar}), and so must also be treated
as massive in Eq.~(\ref{mbpoleMb1S}). We discuss this issue again in
Section~\ref{sectionresults}. 
The functions $\Delta^{\mbox{\tiny
    NLO,NNLO}}_{\mbox{\tiny massive}}$ and $\overline\Delta^{\mbox{\tiny
    NLO}}_{\mbox{\tiny m}}$ are somewhat involved and can be found
in~\cite{Hoang:2000fm,Hoang:1999us}. Charm mass contributions first enter at
order $\norder=2$ in the $\epsilon$ expansion. We will denote order
$\norder=2,3$ results including the charm mass corrections shown in
Eq.~(\ref{mbpoleMb1S}) by $\norder=2_c,3_c$.

\section{Determining $m_c$}

Inclusive $B$ decay spectra have traditionally been computed in terms of the
pole masses $m_b$ and $m_c$. The formul\ae\ used in Ref.~\cite{Bauer:2004ve}
eliminated the pole mass dependence by using Eq.~(\ref{mbmcdiff}) and
Eq.~(\ref{mbpoleMb1S}). The resulting expressions eliminate the $\lqcd$
renormalon from the perturbation series, and have better behaved
perturbative expansions. The fit in 
Ref.~\cite{Bauer:2004ve} used all expressions consistently to order
$\alpha_s^2 \beta_0$, and so used Eq.~(\ref{mbpoleMb1S}), dropping the
$\epsilon^3$ and non-BLM parts of the $\epsilon^2$ terms (which also includes
$\Delta_m^{(2)}$). This paper focuses on the determination of $\mcmu$, and how
the results depend on the order of perturbation theory and the methods that
are used for the analyses. We use the following two methods: \\[1mm]
{\bf Method A:} Take the results of Ref.~\cite{Bauer:2004ve} for $\mbos$,
$\lambda_1$, $\rho_1$ , $\tau_1$ and $\tau_3$ as an input, and 
then determine $\mcmc$ using
Eqs.~(\ref{mbmcdiff},\ref{mcpolemsbar},\ref{mbpoleMb1S}).\\[1mm]
{\bf Method B:} Take the results of Ref.~\cite{Bauer:2004ve} for $\mbos$ and
$m_b-m_c$ as an input, and then determine $\mcmc$ using
Eqs.~(\ref{mcpolemsbar},\ref{mbpoleMb1S}).

 The difference between Methods A and B is whether $m_b-m_c$ or the RHS
side of Eq.~(\ref{mbmcdiff}) is held fixed. Note that for each method the entire
expressions to a given order in $\alpha_s$ are kept, except for the finite charm
mass corrections in Eq.~(\ref{mbpoleMb1S}) as explained below.

Note that in determining $\mcmu$ to order $\epsilon^\norder$,
$\norder={0,1,2,3}$, we will use 
Eqs.~(\ref{mbpoleMb1S}) to order $\epsilon^\norder$, even though the same
expression was only used to order $\alpha_s^2 \beta_0$ in the fit in
Ref.~\cite{Bauer:2004ve}. Thus we break up the problem into two
parts, determining $\mbos$, etc.\ from inclusive $B$ decays, which has already
been studied in detail in Ref.~\cite{Bauer:2004ve}, and then using the results
of the fit to get $\mcmu$.  
In both parts,  the most precise available theoretical predictions are being
used. Since in both parts, the leading ${\cal O}(\lqcd)$ renormalon
contributions are being systematically canceled, correlations of the behavior
of perturbation 
theory in the two parts can be neglected. The errors in each part are
determined separately, and both errors are included in our final error
estimate. The errors in the first part of the problem 
(including the dependence on the order in $\epsilon$) were studied in detail in
Refs.~\cite{Bauer:2002sh,Bauer:2004ve}, and we use the same estimate  here.
 In this work we 
concentrate on the errors in the second part of the problem.

For method B, one substitutes  Eqs.~(\ref{mcpolemsbar},\ref{mbpoleMb1S}) to
write $m_b-m_c$ as 
\begin{eqnarray}
m_b-m_c = f(\alpha_s) \, \mbos - g(\alpha_s)\, \mcmu,
\label{3.1}
\end{eqnarray}
where we have shown explicitly the dependence of $f$ and $g$ on $\alpha_s$,
suppressing the other variables. Both $f(\alpha_s)$ and $g(\alpha_s)$ 
contain an ${\cal O}(\Lambda_{\rm QCD})$ renormalon contribution and have a
badly behaved perturbation series. The  ${\cal O}(\Lambda_{\rm QCD})$
renormalon cancels between the two series, and so is absent in
$m_b-m_c$. To ensure that the series expansion for $m_b-m_c$ is well-behaved,
it is mandatory to truncate  $f(\alpha_s)$ and $g(\alpha_s)$ to the same order
in $\epsilon$, when they are both written in terms of $\alpha_s(\mu)$ \emph{at
  the same scale}. Using different scales, e.g. $\mu=\mbos$ for $f$ and
$\mu=\mcmc$ for $g$ and then truncating at the same order in $\epsilon$ does
not cancel the renormalon, and still leads to a badly 
behaved perturbation series which gives unreliable results.

Solving Eq.~(\ref{3.1}) for $\mcmu$ gives
\begin{eqnarray}
\lefteqn{\mcmum \, = \,} 
\label{3.2} \\[2mm]&&
 { f(\mbos,\alpha_s(\mu),\mu,\overline m(\mu_m),\mu_m) \mbos -
  \left(m_b-m_c\right) \over g\left(\alpha_s(\mu),\mu,\overline
  m(\mu_m),\mu_m\right)}.\nonumber
\end{eqnarray}
We use Eq.~(\ref{3.2}) to determine $\mcmu$ in several different ways. In each
case, we use Eq.~(\ref{3.2}) with $f$ and $g$ expanded to order
$\epsilon^\norder$, with $\norder=0,1,2,2_c,3,3_c$. It is important to note
that expanding to order $\epsilon^\norder$ means that all terms of higher
order \emph{must} be dropped. Since we are dealing with divergent series, one
cannot retain some higher order terms, since they are large. The
cancellations that convert the badly behaved perturbation series for $f$ and
$g$ into a better behaved series for the $\msb$ charm mass only take place
when all terms up to a given order, \emph{and no terms of higher order}, are 
included. This point is discussed in more detail at the end of this section.  

Our final results for $m_c$ are presented in Table~\ref{tab:results}, and are
obtained using the following procedures (in the order given below):  
\begin{enumerate}
\item{} Use $\mu=4.2$~GeV and $\mu_m=4.2$~GeV, to get $\overline m_c(4.2\,
  \text{GeV})$. Eq.~(\ref{3.2}) is expanded out analytically to order
  $\epsilon^\norder$. The terms $\mcmum$ in $f$ and $g$ on the RHS are
  eliminated by iteration resulting in an analytic (but complicated)
  expression. [analytic]
\label{ana42}
\item{} Use $\mu=4.2$~GeV and $\mu_m=4.2$~GeV, to get $\overline m_c(4.2\,
  \text{GeV})$. Eq.~(\ref{3.2}) is solved numerically. In particular, the
  terms $\ln \left[\overline m_c(4.2\,\text{GeV})\right]$ in $f$ and $g$ are
  kept. [numeric] 
\label{num42}
\item{}  Use the result from (\ref{anamc}) for $\mcmc$ and determine $\overline
  m_c(4.2\,\text{GeV})$ using the $(\norder+1)$-loop QCD renormalization group 
  equations. [analytic+RGE] 
\label{anaRGE42}
\item{} Use the result from (\ref{nummc}) for $\mcmc$ and determine $\overline
  m_c(4.2\,\text{GeV})$ using the $(\norder+1)$-loop QCD renormalization group 
  equations. [numeric+RGE] 
\label{numRGE42}
\item{}  Use $\mu=4.2$~GeV and $\mu_m=\mcmc$, to get $\mcmc$. Eq.~(\ref{3.2}) is
  expanded out analytically to order $\epsilon^\norder$. The terms $\mcmc$ in
  $f$ and $g$ are eliminated iteratively, as in (\ref{ana42}). [analytic]
\label{anamc}
\item{}Use $\mu=4.2$~GeV and $\mu_m=\mcmc$, to get $\mcmc$. Eq.~(\ref{3.2}) is
  solved numerically for $\mcmc$. In particular, the terms $\ln \mcmc$
  in $f$ and $g$ are kept. [numeric] 
\label{nummc}
\item{} Use the result from (\ref{ana42}) for $\overline m_c(4.2\,
  \text{GeV})$ and determine $\mcmc$ using the $(\norder+1)$-loop QCD
  renormalization group equations. [analytic+RGE]  
\label{anaRGEmc}
\item{} Use the result from (\ref{num42}) for $\overline m_c(4.2\,\text{GeV})$
  and determine $\mcmc$ using the $(\norder+1)$-loop QCD renormalization group  
  equations. [numeric+RGE] 
\label{numRGEmc}
\end{enumerate} 

For method A the same procedures are followed, except that one uses 
Eq.~(\ref{mbmcdiff}) as the starting point, and all results are (always)
expanded to order $\lqcd^3/m^3$ in addition to the perturbative expansion in 
$\epsilon$ for the functions $f$ and $g$. We note that for the higher order 
$\lqcd/m$ corrections in Eq.~(\ref{mbmcdiff}), we treat the bottom and charm
mass parameter as pole masses and always eliminate the charm mass terms
iteratively in the $\lqcd/m$ expansion. 
For method B, $m_b-m_c$ is determined from Eq.~(\ref{mbmcdiff})
setting the bottom mass terms in the $\lqcd/m_b$ corrections equal to $\mbos$.
Methods A differs from method B through the
dependence of $m_b$ and $m_c$ in the $1/m^2$ and $1/m^3$
terms in Eq.~(\ref{mbmcdiff}) on the order $\norder$. In method B, $m_b-m_c$
is held fixed, whereas 
in method A, it can vary by more than 10~MeV between 
$\norder=0$ and $\norder=3_c$. 

The reference value $\mu=4.2$~GeV is used in the above methods, since that is
the value used in Ref.~\cite{Bauer:2004ve}. The estimate of the theory
uncertainty in Ref.~\cite{Bauer:2004ve} includes half the $\alpha_s^2 \beta_0$
term, and so already includes  any uncertainties due to a different choice of
reference scale. 

\subsection{Renormalon Cancellation}

There are alternative ways of writing Eq.~(\ref{3.2}). For example, one could
also write the mass relations Eqs.~(\ref{mcpolemsbar},\ref{mbpoleMb1S}) in the
inverse form 
\begin{eqnarray}
\mcmu = \tilde g\, m_c , \qquad \mbos = \tilde f\, m_b.
\label{3.3}
\end{eqnarray}
leading to four possible expressions,
\begin{subequations}
\begin{eqnarray}
m_b - m_c &=& f\, \mbos - g\, \mcmu, \\
m_b - m_c &=& {1\over \tilde f}\, \mbos - g\, \mcmu, \\
m_b - m_c &=& f\, \mbos - {1 \over \tilde g}\, \mcmu, \\
m_b - m_c &=& {1\over \tilde f}\, \mbos -{1\over \tilde g}\, \mcmu.
\end{eqnarray}
\end{subequations}
If the functions $f$, $\tilde f$, $g$ and $\tilde g$ each are truncated to
order $\norder$, only the first form represents a well-behaved series in
practice. For the
remaining three cases, it is necessary to reexpand the final expression in
$\epsilon$ and truncate to $\epsilon^\norder$, leading to the first form. For
the analytic procedures described above, all four forms lead then to the same
answer, since the final results are always reexpanded and consistently
truncated to $\epsilon^\norder$. However, for the numeric procedures, where $f$
and $g$ are individually truncated to $\epsilon^\norder$, and then $m_c$
is computed numerically by requiring that the relations analogous to
Eq.~(\ref{3.2}) are satisfied, only
the first form is correct. Solving Eq.~(\ref{3.2}) numerically with any of the
replacements $f\to 1/\tilde f$ or $g \to 1/\tilde g$ can lead to unreliable
results due to badly behaved higher order contributions that remain
uncanceled. This was  
missed in Ref.~\cite{Bauer:2004ve}; the other problem was that $\alpha_s$
was not used at the same renormalization scale $\mu$ for both $f$ and $g$.  

\section{Results and Conclusions}
\label{sectionresults}

\begin{table*}
\caption{\label{tab:results} Results for the $\msb$ charm quark mass in MeV. 
The first error is statistical and includes the uncertainty in $\alpha_s$. The
second  error is the theory uncertainty in the fits to the inclusive spectra
of semileptonic B decays carried out in Ref.~\cite{Bauer:2004ve}.  
The columns are labeled by the order in the $\epsilon$ expansion; 
the subscript $c$ denotes that finite charm quark mass effects (which first
come in at order $\epsilon^2$) have been included in the  1S bottom quark mass
formula in Eq.~(\ref{mbpoleMb1S}).
The upper block uses method A, and the lower block uses method B. The
different methods are discussed in the text.}  
\begin{eqnarray*}
\begin{array}{c|c|c|c|c|c|c|c}
\hline\hline
 & \text{Method}	& 0 & 1 & 2 & 2_c & 3 & 3_c \\
\hline
m_b-m_c  &  & 3401 \pm 7 \pm 11 & 3398 \pm 6 \pm 10 & 3394 \pm 5 \pm 9 & 3394 \pm 5 \pm 9 & 3389 \pm 5 \pm 8 & 3388 \pm 5 \pm 8\\
\overline m_c(4.2\,\mbox{GeV}) & \mbox{analytic} & 1279 \pm 23 \pm 39 & 995 \pm 27 \pm 35 & 897 \pm 33 \pm 33 & 902 \pm 32 \pm 33 & 863 \pm 35 \pm 31 & 880 \pm 33 \pm 31\\
\overline m_c(4.2\,\mbox{GeV}) & \mbox{numeric} & 1279 \pm 23 \pm 39 & 1030 \pm 25 \pm 34 & 929 \pm 29 \pm 33 & 932 \pm 28 \pm 33 & 883 \pm 32 \pm 31 & 895 \pm 30 \pm 32\\
\overline m_c(4.2\,\mbox{GeV}) & \mbox{analytic+RGE} & 1037 \pm 31 \pm 40 & 890 \pm 39 \pm 37 & 845 \pm 43 \pm 35 & 850 \pm 42 \pm 35 & 832 \pm 43 \pm 33 & 850 \pm 41 \pm 33\\
\overline m_c(4.2\,\mbox{GeV}) & \mbox{numeric+RGE} & 1037 \pm 31 \pm 40 & 895 \pm 39 \pm 38 & 851 \pm 42 \pm 36 & 855 \pm 42 \pm 36 & 837 \pm 43 \pm 34 & 851 \pm 41 \pm 34\\
\mcmc & \mbox{analytic} & 1279 \pm 23 \pm 39 & 1213 \pm 21 \pm 35 & 1190 \pm 19 \pm 33 & 1195 \pm 19 \pm 33 & 1184 \pm 18 \pm 30 & 1201 \pm 16 \pm 30\\
\mcmc & \mbox{numeric} & 1279 \pm 23 \pm 39 & 1218 \pm 21 \pm 36 & 1196 \pm 20 \pm 33 & 1200 \pm 19 \pm 33 & 1188 \pm 18 \pm 31 & 1202 \pm 17 \pm 31\\
\mcmc & \mbox{analytic+RGE} & 1521 \pm 21 \pm 37 & 1318 \pm 18 \pm 33 & 1242 \pm 16 \pm 30 & 1246 \pm 16 \pm 30 & 1214 \pm 15 \pm 29 & 1231 \pm 15 \pm 28\\
\mcmc & \mbox{numeric+RGE} & 1521 \pm 21 \pm 37 & 1352 \pm 18 \pm 32 & 1273 \pm 17 \pm 30 & 1276 \pm 17 \pm 30 & 1234 \pm 15 \pm 28 & 1246 \pm 16 \pm 28\\\hline
m_b-m_c  &  & 3401 \pm 7 \pm 11 & 3401 \pm 7 \pm 11 & 3401 \pm 7 \pm 11 & 3401 \pm 7 \pm 11 & 3401 \pm 7 \pm 11 & 3401 \pm 7 \pm 11\\
\overline m_c(4.2\,\mbox{GeV}) & \mbox{analytic} & 1279 \pm 23 \pm 39 & 992 \pm 28 \pm 36 & 891 \pm 34 \pm 35 & 895 \pm 33 \pm 35 & 851 \pm 37 \pm 34 & 868 \pm 35 \pm 34\\
\overline m_c(4.2\,\mbox{GeV}) & \mbox{numeric} & 1279 \pm 23 \pm 39 & 1027 \pm 25 \pm 35 & 923 \pm 29 \pm 34 & 926 \pm 29 \pm 34 & 875 \pm 33 \pm 33 & 886 \pm 32 \pm 34\\
\overline m_c(4.2\,\mbox{GeV}) & \mbox{analytic+RGE} & 1037 \pm 31 \pm 40 & 887 \pm 39 \pm 38 & 838 \pm 44 \pm 37 & 843 \pm 43 \pm 37 & 820 \pm 45 \pm 36 & 837 \pm 43 \pm 36\\
\overline m_c(4.2\,\mbox{GeV}) & \mbox{numeric+RGE} & 1037 \pm 31 \pm 40 & 893 \pm 39 \pm 38 & 845 \pm 43 \pm 37 & 849 \pm 43 \pm 37 & 827 \pm 44 \pm 37 & 841 \pm 43 \pm 37\\
\mcmc & \mbox{analytic} & 1279 \pm 23 \pm 39 & 1210 \pm 21 \pm 36 & 1183 \pm 21 \pm 34 & 1187 \pm 21 \pm 34 & 1172 \pm 20 \pm 33 & 1188 \pm 19 \pm 33\\
\mcmc & \mbox{numeric} & 1279 \pm 23 \pm 39 & 1216 \pm 21 \pm 36 & 1190 \pm 21 \pm 34 & 1194 \pm 20 \pm 35 & 1179 \pm 20 \pm 33 & 1192 \pm 19 \pm 33\\
\mcmc & \mbox{analytic+RGE} & 1521 \pm 21 \pm 37 & 1315 \pm 18 \pm 33 & 1235 \pm 17 \pm 32 & 1240 \pm 17 \pm 32 & 1203 \pm 17 \pm 31 & 1219 \pm 17 \pm 31\\
\mcmc & \mbox{numeric+RGE} & 1521 \pm 21 \pm 37 & 1349 \pm 19 \pm 32 & 1267 \pm 17 \pm 31 & 1270 \pm 18 \pm 31 & 1226 \pm 17 \pm 30 & 1237 \pm 17 \pm 30\\\hline
\hline
\end{array} \nn
\end{eqnarray*}
\end{table*}

The results of our analysis are given in Table~\ref{tab:results}. The input
parameters $\mbos$, $\lambda_1$, $\rho_1$ and $\tau_{1,3}$ and their
covariance matrix needed for our analysis
have been obtained using the 1S-mass fit procedure described in
Ref.~\cite{Bauer:2004ve}, but with the replacement
$\alpha_s(4.2\,\text{GeV})=0.22 \rightarrow 0.2245$  as the central value, to
be consistent with $\alpha_s(M_Z)=0.118$. This makes tiny changes to the input
parameters from the results published in Ref.~\cite{Bauer:2004ve}, e.g.\
$\mbos$ changes by 6~MeV, which is much smaller than its 
$\sim30$~MeV total error. The results for the $b$-quark 1S mass (rounded to $10$~MeV
precision) remains unchanged and reads $\mbos=4.68\pm 0.03$~GeV in full
agreement with the $\Upsilon$ sum rule analyses carried out 
in Refs.~\cite{Hoang:2000fm,Hoang:1999ye}.

For the input parameters used in our analysis,
we have distinguished between experimental uncertainties coming from the data
on inclusive $B$ decay spectra, and the theory uncertainties coming from the
corresponding theoretical predictions at order $\alpha_s^2 \beta_0$ by
carrying out the fits  with theory uncertainties
(as described in Ref.~\cite{Bauer:2004ve}) switched on and switched off.
For our analysis we have also accounted for the uncertainty in
$\alpha_s(4.2\,\text{GeV})$ as an experimental uncertainty, 
including its correlation with $\mbos$, $\lambda_1$, $\rho_1$ and $\tau_{1,3}$. The
corresponding $\alpha_s$ entries of the covariance matrix have been estimated from
the dependence of the best fit for $\mbos$, $\lambda_1$, $\rho_1$ and
$\tau_{1,3}$ on $\alpha_s(4.2\,\text{GeV})$ and assuming that the error on
$\alpha_s(4.2\,\text{GeV})$  is $\pm0.0114$ which corresponds to an error of
$\pm0.003$ for the strong coupling at $\mu=M_Z$. This error has been
conservatively chosen to be  larger than the PDG estimate~\cite{PDG} of
$\pm0.002$ for the error on $\alpha_s(M_Z)$. We find that the uncertainty  in
$\alpha_s(4.2\,\text{GeV})$ has virtually no impact on the total experimental
uncertainty for  $\mcmc$, while the experimental error for $\overline
m_c(4.2\,\text{GeV})$  increases by $15$ to $20$~MeV.

\begin{table*}[]
\caption{\label{tab:linapprox}
Linear formul\ae\ for the $\msb$ charm mass in MeV at order $3_c$ in terms of the
input parameters for the different procedures used in our 
analysis. The input parameters are defined by
$\Delta \alpha_s=(\alpha_s(4.2~\text{GeV})-0.2245)$,
$\Delta m_b=(m_b^{\rm 1S}~\text{GeV}^{-1}-4.681)$,
$\Delta \lambda=(\lambda_1~\text{GeV}^{-2}+0.2426)$,
$\Delta \rho=((\rho_1-\tau_1-\tau_3)~\text{GeV}^{-3}+0.02981)$,
$\delta=((m_b-m_c)~\text{GeV}^{-1}-3.401)$.
The last column shows the maximal deviation in MeV of the linear approximation
from our results over the parameter ranges 
$|\Delta \alpha_s|<0.015$,
$|\Delta m_b|<0.1~\text{GeV}$,
$|\Delta \lambda|<0.1$,
$|\Delta \rho|<0.2$,
$|\delta|<0.05$. 
} 
\begin{eqnarray*}
\begin{array}{c|c|r|c}
\hline\hline
 & \text{Method}	& \multicolumn{1}{c}{3_c} & \text{max. dev.}  \\
\hline
\overline m_c(4.2\,\text{GeV}) & \text{analytic}  & 
 880.7 - 1761 \,\Delta \alpha_s + 159 \,\Delta \lambda +  865 \,\Delta m_b - 59 \,\Delta \rho & 7 \\
\overline m_c(4.2\,\text{GeV}) & \text{numeric}   & 
 895.4 - 1551 \,\Delta \alpha_s + 171 \,\Delta \lambda + 858 \,\Delta m_b - 71 \,\Delta \rho & 7 \\
\overline m_c(4.2\,\text{GeV}) & \text{analytic+RGE}   & 
 849.7 - 2473 \,\Delta \alpha_s + 164 \,\Delta \lambda + 900 \,\Delta m_b - 61 \,\Delta \rho & 7 \\
\overline m_c(4.2\,\text{GeV}) & \text{numeric+RGE}   & 
 850.2 - 2528\,\Delta \alpha_s + 182 \,\Delta \lambda + 913 \,\Delta m_b - 76 \,\Delta \rho & 6 \\
\mcmc  & \text{analytic}	                  & 
 1202.0 + 272 \,\Delta \alpha_s + 162 \,\Delta \lambda + 886 \,\Delta m_b - 60 \,\Delta \rho & 8  \\
\mcmc & \text{numeric}                            & 
 1202.6 + 217 \,\Delta \alpha_s + 179 \,\Delta \lambda + 899 \,\Delta m_b - 75 \,\Delta \rho & 7 \\
\mcmc & \text{analytic+RGE}                        	  & 
 1232.5 + 937 \,\Delta \alpha_s + 156 \,\Delta \lambda + 850 \,\Delta m_b - 58 \,\Delta \rho &  9 \\ 
\mcmc & \text{numeric+RGE}                            & 
 1247.5 + 1126 \,\Delta \alpha_s + 168 \,\Delta \lambda + 842 \,\Delta m_b - 70 \,\Delta \rho & 8 \\
\hline
\overline m_c(4.2\,\text{GeV}) & \text{analytic}  & 
 868.0 - 1920 \,\Delta \alpha_s + 836 \,\Delta m_b - 808 \,\delta & 4 \\
\overline m_c(4.2\,\text{GeV}) & \text{numeric}   & 
 886.6 - 1635 \,\Delta \alpha_s + 823 \,\Delta m_b - 798 \,\delta & 5 \\
\overline m_c(4.2\,\text{GeV}) & \text{analytic+RGE}   & 
 836.6 - 2650 \,\Delta \alpha_s + 869 \,\Delta m_b - 840 \,\delta & 2 \\
\overline m_c(4.2\,\text{GeV}) & \text{numeric+RGE}   & 
 840.5 - 2632 \,\Delta \alpha_s + 877 \,\Delta m_b - 850 \,\delta & 2 \\
\mcmc  & \text{analytic}	                  & 
 1188.9 + 110\,\Delta \alpha_s + 858 \,\Delta m_b - 829 \,\delta & 3  \\
\mcmc & \text{numeric}                            & 
 1192.7 + 124 \,\Delta \alpha_s + 865 \,\Delta m_b - 838 \,\delta & 3 \\
\mcmc & \text{analytic+RGE}                        	  & 
 1219.8 + 793 \,\Delta \alpha_s + 822 \,\Delta m_b - 795 \,\delta & 7 \\
\mcmc & \text{numeric+RGE}                            & 
 1238.1 + 1052 \,\Delta \alpha_s + 809 \,\Delta m_b - 784 \,\delta & 8 \\
\hline\hline
\end{array} \nn
\end{eqnarray*}
\end{table*}

The various results shown in Table~\ref{tab:results} are stable, and reflect a
very good perturbative behavior in each case. For each entry in the table the
first error is statistical accounting for the uncertainties and correlations
of the input parameters $\mbos$, $\lambda_1$, $\rho_1$, $\tau_{1,3}$ and
$\alpha_s(4.2\,\text{GeV})$ due to uncertainties in the experimental
data used in the analysis of Ref.~\cite{Bauer:2004ve}, and the second error
accounts for the theory uncertainties  
in the predictions used in Ref.~\cite{Bauer:2004ve}.
The results show some interesting features that are worthy of comment:
It is conspicuous that the results for $\mcmc$ obtained directly in a single
step appear more stable than those
obtained by scaling $\overline m_c(4.2\,\text{GeV})$ down to the charm
mass. For example for method A, the two results for $\mcmc$ based on the analytic  
determination read 
$(1279-66\epsilon-18 \epsilon_c^2+6\epsilon_c^3)$~MeV [analytic] and
$(1521-203\epsilon-72 \epsilon_c^2-15\epsilon_c^3)$~MeV [analytic+RGE].
Both series show very good convergence properties. On the other hand, the
differences between the central values for the two series at order
(1,$\epsilon$,$\epsilon_c^2$,$\epsilon_c^3$) are $(242,105,51,30)$~MeV. These values
are much larger than one might guess by using a naive error estimate based on
the good behavior of each individual series. This is a feature that appears
common 
to the behavior of perturbation theory at the (relatively low) charm mass
scale~\cite{Hoang:2004xm}. Since there is no a-priori  reason to consider 
any single one of the series to be more reliable than the other, we use as an
estimate of the theoretical uncertainty in our analysis the spread of central
values for the different methods.  

It is instructive to compare the results at order $2,3$ and $2_c,3_c$. While
accounting for the charm mass corrections in the relation of the bottom pole
and 1S masses, Eq.~(\ref{mbpoleMb1S}) leads only to a small difference of at most
$5$~MeV at order $2$ and $2_c$ for $\mcmc$, the difference is already as large
as the experimental uncertainties of around $15$~MeV at order $3$ and
$3_c$. This behavior arises because treating the charm as massless in the
function $f$ of Eq.~(\ref{mbpoleMb1S}) leads to a different 
${\cal O}(\Lambda_{\rm QCD})$-renormalon behavior in the bottom
pole-1S mass series than in the charm pole-$\msb$ mass series where the
charm mass effects are automatically included. Thus, as explained in
Section~\ref{sectionrelations}, at order $2$ and $3$ the ${\cal
  O}(\Lambda_{\rm QCD})$ renormalon contributions contained in $f$ and $g$ do
not fully cancel. Therefore, to avoid this source of perturbative ambiguity it
is mandatory to account for the charm mass effects in the bottom pole-1S mass
relation, and only the orders $2_c$ and $3_c$ should be considered for the
analysis. 

To determine our final result for $\mcmc$ we use the order $\epsilon_c^3$
results. The results obtained using
method A and B are very similar. For the central value and the
error estimate we therefore only use the numbers from method A.  
As the central value we use the average of the largest and smallest
numbers. For the experimental uncertainty we take the largest experimental
error of $17$~MeV. For the
theory uncertainty, we linearly add the error originating from the theory error 
of the input parameters $\mbos$, $\lambda_1$, $\rho_1$ and $\tau_{1,3}$ of
$31$~MeV, and the perturbative uncertainty of our analysis. For the latter we
take half the 
range of the central values we have obtained, which is $23$~MeV.  This gives
\begin{eqnarray}
\mcmc & = & 1224 \, \pm \, 17_{\text{exp}} \, \pm \, 31_B \pm 23_{m_c}~\text{MeV}
\,,
\end{eqnarray}
with the subscripts denoting the sources of the various uncertainties.
We add the
two theoretical uncertainties linearly, to be conservative. Moreover, this
more conservative treatment should 
also account for the uncertainty arising from the unknown order
$\alpha_s\Lambda_{\rm QCD}^2/m$ corrections in Eq.~(\ref{mbmcdiff}). The latter
uncertainty is reflected by the variation of $m_b-m_c$ with the order $\norder$,
and is about $10$~MeV. Thus our final result is
\begin{eqnarray}
\mcmc & = & 1224 \, \pm \, 17 \, \pm \, 54~\text{MeV}
\,.
\end{eqnarray}
The first error is the experimental 1-$\sigma$ error  and the second is the
theoretical error. Note that the theoretical error should not be interpreted
as a 1-$\sigma$ error, since it is not statistical. Rather it represents
the range in which we believe that the true charm mass is located with
high probability (much greater than the 67\% probability of a 1-$\sigma$
interval). Our result is consistent with the most recent result for
$\mcmc$ from low-$n$ $e^+e^-$ moment sum rules~\cite{Hoang:2004xm}. However, 
our result is somewhat lower than the most recent results obtained from
lattice QCD~\cite{Rolf:2002gu,deDivitiis:2003iy}, which were
obtained in the quenched approximation.  

Linear approximation formul\ae\ showing the  
dependence of our results on the input parameters $\mbos$, $\lambda_1$,
$\rho_1$, $\tau_{1,3}$, $\alpha_s(4.2~\text{GeV})$ and $m_b-m_c$ for methods A
and B at order $\epsilon_c^3$ are displayed in Table~\ref{tab:linapprox}.

In our analysis we have also determined $\overline m_c(4.2\,\text{GeV})$ using
the same procedure as used for $\mcmc$,
\begin{eqnarray}
\overline m_c(4.2\,\text{GeV})\!\! & = &\!\! 873 \, \pm \, 41_{\text{exp}} \, \pm \, 34_B
\pm 23_{m_c}~\text{MeV}\,.\
\end{eqnarray}
Our final result is 
\begin{eqnarray}
\overline m_c(4.2\,\text{GeV}) & = & 873 \, \pm \, 41 \, \pm \, 57~\text{MeV}
\,.
\end{eqnarray}
While the theoretical 
uncertainties are comparable to the corresponding uncertainties for $\mcmc$, we
find that the experimental errors are $15$ to $20$ MeV larger. As
mentioned before,  this increase originates from the
uncertainties in $\alpha_s$ which have a rather large impact on the
uncertainty in $\overline m_c(4.2\,\text{GeV})$, but almost none for $\mcmc$.
The origin of this behavior can be clearly seen from the linear approximation
formulae shown in Table~\ref{tab:linapprox}: the values of $\mcmc$ and 
$\overline m_c(4.2\,\text{GeV})$ both increase for larger input values
$\mbos$; on the other hand, while the results for $\mcmc$
increase with $\alpha_s$, the values $\overline m_c(4.2\,\text{GeV})$
decrease. 
The fits of the semileptonic inclusive B decay spectra show that
$\mbos$ becomes smaller for larger $\alpha_s$ values, and so the $\Delta m_b$
term tends to cancel the $\Delta \alpha_s$ term for $\mcmc$, but comes in with
the same sign for $\overline m_c(4.2\,\text{GeV})$.

Based on our error analysis it is straightforward to discuss how the
uncertainties in our charm mass analysis might be further reduced in the
future. More precise experimental data on the semileptonic inclusive $B$ decay
spectra can not reduce the error significantly because they constitute
only a relative small contribution to the total error. In this respect our
analysis is similar to sum rule analyses based on the $e^+e^-$ R-ratio and
lattice QCD where theory uncertainties dominate. A more precise input value for
$\alpha_s$ will not affect the uncertainties of $\mcmc$, but can reduce
substantially the experimental error of $\overline m_c(4.2\,\text{GeV})$. 
A full ${\cal O}(\alpha_s^2)$ analysis of the semileptonic inclusive $B$ decay
spectra would reduce the theory uncertainty of the input parameters
$\mbos$, $\lambda_1$, $\rho_1$ and $\tau_{1,3}$ of our charm mass analysis.
Moreover, a determination of ${\cal O}(\alpha_s\lqcd^2/m^2)$ corrections 
would reduce the scheme dependence of the $\lambda_{\rm QCD}^2/m$
contributions in the OPE for the $B$-$D$ meson mass difference. 
We believe that these  computations are achieveable with present technology 
and could reduce the theory errors shown in Table~\ref{tab:results} by at
least a factor of two.
Finally, a computation of the full order $\epsilon_c^4$ corrections to the
bottom-charm pole mass difference as a function of the $\msb$ charm mass and
the bottom 1S mass could further reduce the theoretical uncertainty in
Eq.~(\ref{3.2}). Such work would allow one to reduce the uncertainty of
$\mcmc$, with a conservative estimation of theory errors,  to well below the
$50$~MeV level.

\begin{acknowledgments} 
AM would like to thank the Alexander von Humboldt foundation for support.
\end{acknowledgments}



\begin{thebibliography}{}

\bibitem{Buras:2005gr}
A.~J.~Buras, M.~Gorbahn, U.~Haisch and U.~Nierste,
arXiv:hep-ph/0508165.
 
\bibitem{Brambilla:2004wf}
  N.~Brambilla {\it et al.},
  arXiv:hep-ph/0412158.

\bibitem{Kuhn:2001dm}
  J.~H.~Kuhn and M.~Steinhauser,
  Nucl.\ Phys.\ B {\bf 619}, 588 (2001)
  [Erratum-ibid.\ B {\bf 640}, 415 (2002)]
  [arXiv:hep-ph/0109084].

\bibitem{Hoang:2004xm}
  A.~H.~Hoang and M.~Jamin,
  Phys.\ Lett.\ B {\bf 594}, 127 (2004)
  [arXiv:hep-ph/0403083].


\bibitem{Rolf:2002gu}
  J.~Rolf and S.~Sint  [ALPHA Collaboration],
  JHEP {\bf 0212}, 007 (2002)
  [arXiv:hep-ph/0209255].

\bibitem{deDivitiis:2003iy}
  G.~M.~de Divitiis, M.~Guagnelli, R.~Petronzio, N.~Tantalo and F.~Palombi,
  Nucl.\ Phys.\ B {\bf 675}, 309 (2003)
  [arXiv:hep-lat/0305018].


\bibitem{Bauer:2004ve}
  C.~W.~Bauer, Z.~Ligeti, M.~Luke, A.~V.~Manohar and M.~Trott,
  Phys.\ Rev.\ D {\bf 70}, 094017 (2004)
  [arXiv:hep-ph/0408002].

\bibitem{Bauer:2002sh}
  C.~W.~Bauer, Z.~Ligeti, M.~Luke and A.~V.~Manohar,
  Phys.\ Rev.\ D {\bf 67}, 054012 (2003)
  [arXiv:hep-ph/0210027].


\bibitem{Gambino:2004qm}
P.~Gambino and N.~Uraltsev,
Eur.\ Phys.\ J.\ C {\bf 34}, 181 (2004)
[arXiv:hep-ph/0401063].

\bibitem{Aubert:2004aw}
B.~Aubert {\it et al.}  [BABAR Collaboration],
Phys.\ Rev.\ Lett.\  {\bf 93}, 011803 (2004)
[arXiv:hep-ex/0404017].

\bibitem{bigi}
  I.~I.~Y.~Bigi, M.~A.~Shifman, N.~G.~Uraltsev and A.~I.~Vainshtein,
  Phys.\ Rev.\ D {\bf 50} (1994) 2234.

\bibitem{beneke}
  M.~Beneke and V.~M.~Braun,
  Nucl.\ Phys.\ B {\bf 426} (1994) 301.

\bibitem{Melnikov:2000qh}
  K.~Melnikov and T.~v.~Ritbergen,
  Phys.\ Lett.\ B {\bf 482}, 99 (2000)
  [arXiv:hep-ph/9912391];
see also
  K.~G.~Chetyrkin and M.~Steinhauser,
  Nucl.\ Phys.\ B {\bf 573}, 617 (2000)
  [arXiv:hep-ph/9911434].

\bibitem{Hoang:1998ng}
  A.~H.~Hoang, Z.~Ligeti and A.~V.~Manohar,
  Phys.\ Rev.\ Lett.\  {\bf 82}, 277 (1999)
  [arXiv:hep-ph/9809423].

\bibitem{Hoang:1998hm}
  A.~H.~Hoang, Z.~Ligeti and A.~V.~Manohar,
  Phys.\ Rev.\ D {\bf 59}, 074017 (1999)
  [arXiv:hep-ph/9811239].


\bibitem{Hoang:2000fm}
  A.~H.~Hoang,
  arXiv:hep-ph/0008102.

\bibitem{Hoang:1999ye}
  A.~H.~Hoang,
  Phys.\ Rev.\ D {\bf 61}, 034005 (2000)
  [arXiv:hep-ph/9905550].

\bibitem{Hoang:1999us}
  A.~H.~Hoang and A.~V.~Manohar,
  Phys.\ Lett.\ B {\bf 483}, 94 (2000)
  [arXiv:hep-ph/9911461].



\bibitem{PDG}
S.~Eidelman {\it et al.}  [Particle Data Group Collaboration],
Phys.\ Lett.\ B {\bf 592}, 1 (2004).




\end{thebibliography}
\end{document}